\documentclass[prl,showpacs,showkeys,twocolumn,floatfix]{revtex4} 
\usepackage[]{epsfig,dcolumn}
\begin{document}
\title{The Evidence for a  Pentaquark Signal and Kinematic Reflections}
\author{A.~R.~Dzierba}
\author{D.~Krop}
\author{M.~Swat}
\author{S.~Teige}
\affiliation{Department of Physics, Indiana University, Bloomington, IN 47405}
\author{A.~P.~Szczepaniak}
\affiliation{Departemnt of Physics and Nuclear Theory Center, 
 Indiana University, Bloomington, IN 47405}
\date{\today}

\begin{abstract}
Several recent experiments have reported evidence for
 a narrow baryon resonance with positive strangeness ($\Theta^+$)
 at a mass of 1.54 GeV/$c^2$.  Baryons with $S=+1$ cannot
 be conventional $qqq$ states and the reports have thus  generated much 
 theoretical speculation about the nature of  possible $S=+1$ baryons,
  including a 5-quark, or pentaquark, interpretation.  We show that  
 narrow enhancements in the $K^+n$ effective mass spectrum can be generated 
 as  kinematic reflections resulting from the decay of mesons, 
 such as the $f_2(1275)$, the  $a_2(1320)$ and the $\rho_3(1690)$.
\end{abstract}

\maketitle

Several experiments have recently claimed evidence for a so-called 
 $\Theta^+$, a positive strangeness baryon resonance  with a very narrow 
 width  \cite{spring8,clas,saphir,diana}.  Baryons with $S=+1$ cannot be 
simple $qqq$ states and indeed the failure to find  $S=+1$ baryons
in the 1960's (they were then called $Z^*$'s) gave credence to the
original quark model.  QCD can accommodate  $S=+1$ baryons as
more complicated structures such a five-quark states known as
\emph{pentaquarks}.  Indeed, the recent experimental claims have generated
 much theoretical interest and speculation about the nature of the $\Theta^+$
and the possibility of searching for other states that may be members of a
larger multiplet.  

Guided by  the principle of parsimony and by the fact that earlier searches
 did not uncover $S=+1$ baryon resonances, we
 examine more conventional mechanisms that could account for narrow 
 structures in meson-baryon effective mass combinations with $S=+1$.
 That is the focus of this paper.  We examine the effect of the constrained
 kinematics characterizing the experiments reporting the $\Theta^+$ and the
possibility that details of the decay angular distribution of known meson 
 resonances could account for  enhancements in effective mass, 
 \emph{e.g.} $K^+n$, distributions.

It is interesting to note that in an earlier search \cite{anderson} for a  
 $Z^*$ in the reaction  $\pi^- p \to K^- Z^*$ at incident 
 $\pi^-$ momenta of 6 and 8~GeV/$c$ found enhancements in the missing mass 
 recoiling off of the $K^-$ near 1.5 and 1.9~GeV/$c^2$.
 The authors explained the enhancements as being kinematic reflections 
 of the $f_2(1275)$, the  $a_2(1320)$ and the $\rho_3(1690)$.  
 It was also observed that the  peak position of the enhancements 
 changed with changing beam momentum supporting their interpretation of the 
 enhancements.


 Three of the experiments \cite{spring8,clas,saphir}
 claiming the $\Theta^+$ used incident photons scattering off of
 $p$ \cite{saphir}, $d$ \cite{clas} and $^{12}C$ \cite{spring8} 
 targets.  Two of these experiments produced bremsstrahlung photons
 beams with electron energies of 2.47 and 3.1~GeV \cite{clas} and
 2.8 GeV \cite{saphir} (the photon spectrum extends up to 
 95\% of the electron energy)
  and the other used Compton backscattered
 photon beams starting with electron energies above 1.5 GeV and less
 than 8 GeV.  One of the experiments \cite{diana} used a low energy $K^+$ beam 
 scattered off of a $Xe$ target. 
 
 \begin{figure}
\centerline{\epsfig{file=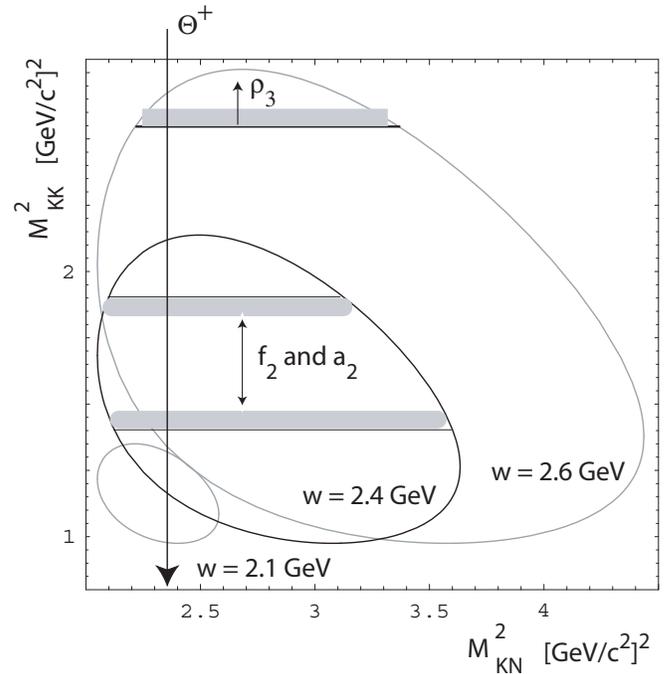,width=\columnwidth}}
\caption{ Boundaries of the $m_{KK}^2$ versus $m_{KN}^2$ Dalitz plot for
three different values of $w$, the energy available to the $K \bar K N$ 
 system, 2.1, 2.4 and 2.6 GeV.  For the data of ref. \cite{clas}, 
 the observed distribution in $w$ rises from 2.1 GeV, peaks at 2.4 and 
 falls to zero near 2.6 GeV.
 Horizontal lines denote the region spanned by the 
 $f_2$ and $a_2$ mesons defined by their half-widths
 and the region of the $\rho_3$ starting with its central mass less
 its half-width.  The vertical line denotes the 
 square of the $\Theta$ mass.}\label{dalitz}
\end{figure}

 Figure \ref{dalitz} shows the kinematic boundaries of the $m_{KK}^2$ 
 versus $m_{KN}^2$ Dalitz plot for three different values of $w$, 
 the energy available to the $K \bar K N$ system,
 2.1, 2.4 and 2.6 GeV.  The observed value of $w$ is determined by the beam 
momentum, the Fermi momentum of the target and the cuts used to define
 the data sample. For the data of ref. \cite{clas}, the observed distribution
 in $w$ rises from 2.1 GeV, peaks at 2.4 and falls to zero near 2.6 GeV.
   The other experiments reporting the $\Theta^+$ have
 similar ranges of energies available to the $K \bar K N$ system.
 Horizontal lines denote the region spanned by the 
 $f_2$ and $a_2$ mesons defined by their half-widths
 and the region of the $\rho_3$ starting with its central mass less
 its half-width. The vertical line denotes the  square of the $\Theta$ mass.  
 As the energetically allowed area of the Dalitz plot grows from smallest 
 to largest the intersection of the $f_2$, then $a_2$ and finally 
 $\rho_3$ bands with the boundary cluster about a value of $m_{KN}$ near 
 the $\Theta$ mass.

\begin{table}
\begin{tabular*}{3.5in}{@{\extracolsep{0.1in}}lcclr}
Reaction & Beam energy & Cross Section& Ref\\
& GeV & $\mu$b \\ \hline \hline

$\gamma {\rm{p}} \rightarrow {\rm{f_2}} {\rm{p}} $ & 2.3-2.6   & $1.3   \pm 0.37 $&\cite{stru}  \\ 
$\gamma {\rm{p}} \rightarrow {\rm{f_2}} {\rm{p}} $ & 2.6-3.25 & $0.39 \pm 0.13 $&\cite{stru} \\
$\gamma {\rm{p}} \rightarrow {\rm{f_2}} {\rm{p}} $ & 3.25-4.0 & $0.19 \pm 0.06 $&\cite{stru} \\
$\gamma {\rm{p}} \rightarrow {\rm{f_2}} {\rm{p}} $ & 4.0-6.3   & $0.1    \pm 0.1 $&\cite{stru} \\ 

$\gamma {\rm{p}} \rightarrow {\rm{a_2^+}} {\rm{n}} $ & $ 4.2 \pm 0.5$   & $ 1.14 \pm 0.43 $&\cite{Eisenberg} \\ 
$\gamma {\rm{p}} \rightarrow {\rm{a_2^+}} {\rm{n}} $ & $ 5.25 \pm 0.55$   & $ 0.85 \pm 0.43 $& \cite{Eisenberg}\\ 
$\gamma {\rm{p}} \rightarrow {\rm{a_2^+}} {\rm{n}} $ & $ 7.5 \pm 0.7$   & $ 0.43 \pm 0.43 $&\cite{Eisenberg} \\

$\gamma {\rm{p}} \rightarrow {\rm{K^+}} {\rm{K^-}} {\rm{p}} $ & 2.8   & $1.0    \pm 0.1 $& \cite{Ballam} \\ 
$\gamma {\rm{p}} \rightarrow {\rm{K^+}} {\rm{K^-}} {\rm{p}} $ & 4.7   & $0.7    \pm 0.1 $&\cite{Ballam} \\ \hline

\end{tabular*}
 \caption{Photoproduction cross sections for the $f_2(1275)$ and $a_2(1320)$ 
 resonances and the $K^+ K^-$ final state.}
\label{sigmas}
\end{table}

 Because of the constrained kinematics, the projections of the Dalitz plot 
 onto the $m_{KN}$ axis can be strongly influenced by the decays of 
 $K \bar K$ resonances. Consider the transformation from the
 $K \bar K N$ system in its center-of-mass to the helicity frame of the 
 $K \bar K$ system (center of mass of the $K \bar K$ with the $K \bar K$ 
 line-of-flight defining the z-axis). In this frame the decay angular 
 distribution of the $K \bar K$ is described by $|Y^M_L(\theta,\phi)|^2$ 
 where $L=1$ for the $\phi(1020)$,  $L=2$ for the $f_2(1275)$ and the 
 $a_2(1320)$  and  $L=3$ for the $\rho_3(1690)$.  Several of these mesons 
 are known to have photoproduction cross sections
 approaching the total cross section for $K^+ K^-$ (see Table \ref{sigmas}). 
 All of them have well established decays into $K^+ K^-$ with branching 
 fractions of several percent making plausible the argument that 
 production and subsequent decay of these mesons can produce contributions 
 to observed distributions of the same size as the observed $\Theta^+$. 
 As the spin ($L$) of the mesons increases, the greater the alignment
  of the decay products parallel to or anti-parallel to the line of flight 
 of the meson -- or alternatively -- parallel or anti-parallel to the 
 nucleon ($N$).  In fact, the distribution in the decay angles
 correlates with the $KN$ mass distribution.  

Experiments (\emph{e.g.} ref. \cite{clas}) have removed events with
$m_{K \bar K}$~$<$~1.07~GeV/$c^2$ to avoid the strongly-produced
$\phi(1020)$ but not the higher mass and higher spin resonances.  

\begin{figure}  
\centerline{\epsfig{file=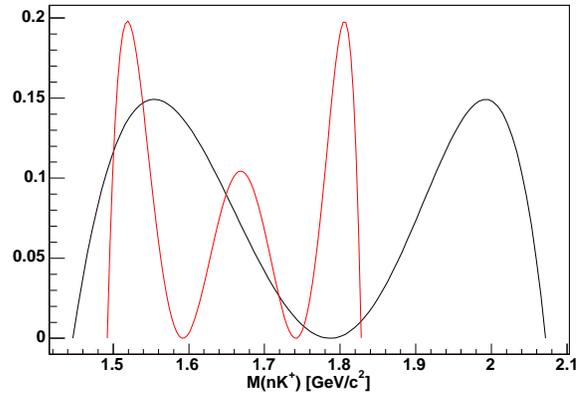,width=\columnwidth}}
\caption{ The $m_{KN}$ mass distribution for a fixed $m_{KKN}$ mass of 
 2.6~GeV/$c^2$.
 The two-peaked curve assumes $m_{KK} = m(a_2)$
 with $|Y^{\pm 1}_2|^2$ and  and the three-peaked curve
 assumes $m_{KK} = m(\rho_3) - \Gamma_{\rho}/2$
 with $|Y^{\pm 1}_3|^2$ for the decay angular distribution.}\label{reflecs}
\end{figure}

Figure \ref{reflecs} shows the  $m_{KN}$ distribution obtained from simply
 assuming production of an $a_2(1320)$ with no-helicity-flip {\it i.e} 
 with a decay angular distribution described by 
 $|Y^{\pm 1}_2(\theta,\phi)|^2$ (two-peaked curve) and the 
 $\rho_3(1690)$ described by $|Y^{\pm 1}_3(\theta,\phi)|^2$ 
 (three-peaked curve).  For the plot of
 this figure the mass of the $K \bar K N$  system was assumed to be 
 2.6~GeV/$c^2$.   The plot illustrates
 how sharp peaks can be generated and how the $m_{KN}$ distribution
 is reflected in the form of the decay angular distribution.

The enhancements shown in Figure \ref{reflecs} will be broadened after 
 integrating over $K \bar K N$ energy -- variations due to the spread 
 in beam momentum and  Fermi motion -- and the $K^+ K^-$ mass. 
To study these broadening effects, we focus on the data of ref. \cite{clas} 
 for $\gamma d \to K^+K^- n p$.  We assume that the $K^+K^-$ results 
 from a coherent production of resonances $X=a_2,f_2,\rho_3$. 
 In addition, as in ref.~\cite{clas}, we assume that the reaction takes 
 place off of the neutron with the proton in the deuteron acting as a 
 spectator, $\gamma n \to X n \to K^+ K^- n$. At the energies of ref. 
 \cite{clas}, ranging from $E_\gamma = 1.51\mbox{ GeV}$ to 
 $E_\gamma = 2.96\mbox{ GeV}$, meson production can be effectively 
 described by $t$-channel, Regge exchange amplitudes, 
\begin{equation}
A(\gamma n \to K^+K^- n) = P(\gamma n \to X n) D(X \to K^+ K^-)
\end{equation} 
where
\begin{equation}
P(\gamma n \to K^+ K^- n) = s^{\alpha(t)} R(t)  
|t-t_{\min}|^{n/2}, \label{prod} 
\end{equation}
and
\begin{equation}
D(X\to K^+ K^-) = {  { k^L Y_{LM}(\Omega_H) } \over 
 {( m_{KK}^2 - m_X^2) + im_X \Gamma(m_{KK})}}. \label{decay}
\end{equation}
 $a_2$ photoproduction has been studied 
 previously~\cite{Eisenberg, Condo, ASAA} and can be well described by a 
 $\pi$ exchange corresponding to $\alpha(t) = 0.9 (t - m_\pi^2) 
 \mbox{ GeV}^{-2}$ and $R(t) = \exp(bt)$ with $b = 10 -30 \mbox{ GeV}^{-2}$ 
 with $b = 3 \mbox{ GeV}^{-2}$ for low-$t$ ($|t| <  0.2 \mbox{ GeV}^2$) 
 and high-$t$, {\it respectively}. For simplicity we assume the same
 form of energy and momentum transfer dependence for production of
 the other resonances. The helicity structure of the production
 amplitude is given by last factor in Eq.~(\ref{prod}), with $n$ equal
 to the net helicity flip. For nucleon-flip pion exchange, assuming
 Regge factorization, we have $n = 1 + |1 - M|$ where $M$ is the
 helicity of the resonance. In the decay amplitude of
 Eq.~{\ref{decay}), $L$ stands for the spin of the resonance, $k$ is
 the resonance breakup momentum, $k = \sqrt{m_X^2/4 - m_K^2}$,
 $\Omega_H$ represents the helicity angles of the $K^+$ and $m_X$ and 
 $\Gamma(m)$ are resonance's mass and energy-dependent width, 
 respectively. To compare with the data of ref. \cite{clas} we convolve
 the  theoretical cross section calculated from amplitudes given above
 with the bremsstrahlung photon spectrum and the neutron Fermi momentum 
 distribution in a deuteron. The $m_{KN}$ spectrum is then obtained by 
 integrating over all remaining kinematic variables subject to the 
 requirement that $m_{KK} > 1.07 \mbox{ GeV}$ and is shown in
 Figure~3. The theoretical
 spectrum  depends on (complex) resonance strengths, in addition to a 
 coherent $S$ and $P$ wave production. These parameters are obtained
 from  fitting the  $m_{KN}$ distribution from ref.~\cite{clas}.  
 The low-angular momentum waves are expected due to residual $K{\bar
 K}$  interactions which, for example, are responsible for the
 $f_0(980)$  meson. At higher $K^+K^-$ mass one expects contributions
 from  the $f_0(1370)$ and higher scalar resonances. Finally, the presence of
 the  $P$-wave above the $\phi(1020)$ has been established in other 
 photoproduction experiments,~\cite{Fries,Barber,ASLL}. The
 theoretical  $K^+K^-$ distribution fixed by the $m_{KN}$ data is then 
 compared with the experimental $K^+K^-$ distribution from  
 ref.~\cite{clas} and is shwon in Figure~4. 
 The agreement is very good considering that the
 $P$-wave is constrained outside the $\phi(1020)$ region.

\begin{figure} 
\includegraphics[width=3.in]{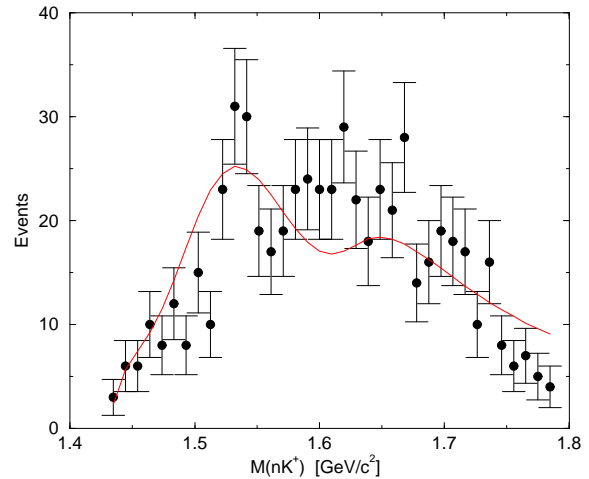}
\caption{\label{fig3} The calculated (solid line) $m_{KN}$
  distribution, as described in the text, compared with the data 
 from~\cite{clas} }   
\end{figure}

\begin{figure}
\includegraphics[width=3.5in]{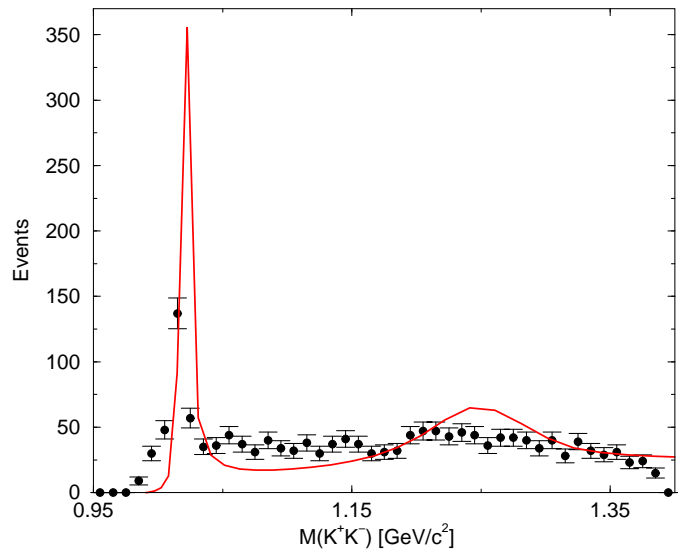} 
\caption{\label{fig4} The calculated (solid line) $m_{KK}$
  distribution,, as described in the text, compared with the data 
 from~\cite{clas} }  
\end{figure} 

It is clear that a enhancement at $m_{KN}$ close to the purported $\Theta^+$ mass, due to peaks in the $K^+K^-$ angular distribution, remains after taking into account the cross section and $s$- distributions. 


We have shown that kinematic reflections due to meson resonances could well
account for the enhancement observed in the $K^+ n$ effective mass 
distribution at the mass of the purported  $\Theta^+$.  Further experimental
studies will be required with higher statistics, varying the incident
beam momentum and  establishing the spin and parity before claiming
solid evidence for a $S=+1$ baryon resonance.

The authors acknowledge useful discussions with Curtis Meyer and Elton Smith.
  This work was supported in part by the  National Science Foundation
  and the U. S. Department of Energy.

\pagebreak

\end{document}